\documentclass[aps,prl,twocolumn,groupedaddress,showpacs]{revtex4}
\usepackage{graphicx}
\graphicspath{{figure file fold path}}
\usepackage[dvipsnames,usenames]{color}
\usepackage{color}
\usepackage{amsmath}
\usepackage{float}
\usepackage{amssymb}
\usepackage{amsthm}
\usepackage{mathrsfs}

\emergencystretch=\maxdimen
\begin{document}

\title{Polarization-Multiplexed Chaotic LiDAR Based on a VCSEL with Delayed Orthogonal Feedback}
\author{Tao Wang$^{1}$, Zhibo Li$^{1}$, Hui Shen$^{1}$, Yixing Ma$^{1}$, Yiheng Li$^{1}$, Shuiying Xiang$^{1}$, Stephane Baland$^{2}$, and Yue Hao$^{3}$}

\affiliation{$^1$State Key Laboratory of Integrated Service Networks, School of Telecommunications Engineering, Xidian University, Xi'an 710071, China}
\affiliation{$^2$Universit\'e C\^ ote d’Azur, Institute de Physique de Nice, UMR 7010 CNRS, Nice, 06200, France}
\affiliation{$^3$State Key Discipline Laboratory of Wide Bandgap Semiconductor Technology, School of Microelectronics, Xidian University, Xi’an 710071, China}

\date{\today}

\begin{abstract}
Light detection and ranging (LiDAR) systems are pivotal for precise distance and velocity measurement, yet widespread deployment requires solutions that balance their performance, robustness, and simplicity. Here, we propose a novel chaotic LiDAR system based on a semiconductor vertical-cavity surface-emitting laser (VCSEL) with delayed orthogonal polarization feedback. By exploiting the intrinsic competition between the transverse electric (TE) and transverse magnetic (TM) modes, the system generates a polarization-multiplexed dynamics: a chaotic TM mode serves as the reference, while a feedback-modulated TE mode probes the target. This all-in-one source eliminates the need for external optical modulators or complex coherent detection. The system's dynamics is finely tunable via a half-wave ($\lambda$/2) plate in the feedback loop and the laser injection current, enabling real-time optimization of the cross-correlation signal-to-noise ratio. Experimental results demonstrate precise linear ranging with a resolution of approximately 1.2 cm. Furthermore, the system exhibits strong inherent resistance to external optical interference, maintaining accurate ranging even in the presence of a secondary laser source. This compact, tunable, and interference-resilient platform offers a promising pathway toward low-cost, high-performance LiDAR for applications in autonomous navigation, robotics, and industrial metrology.  
\end{abstract}


\maketitle 

\section{Introduction}
As an advanced environmental perception technology, LiDAR plays a significant role in various fields~\cite{riemensberger2020} such as autonomous driving~\cite{ibanez2025lidar}, robotics~\cite{malladi2025}, and airborne remote sensing~\cite{mercier2025}. Conventional LiDAR systems often employ pulsed lasers as the light source, determining target distance through the time-of-flight (TOF) of back-scattered optical pulses~\cite{lin2025mid, irwin2025light}. This method is well-suited for long-range applications but requires careful management of high peak powers to comply with laser safety standards~\cite{ma2024review}. In contrast, frequency-modulated or amplitude-modulated continuous-wave (FMCW/AMCW) LiDAR systems emit a continuously modulated laser beam and derive distance information from the frequency difference between the transmitted and reflected signals~\cite{li2022high, lee2023automatic}. While enabling high precision and sensitivity, such systems demand ultra-narrow linewidth tunable lasers and stable coherent detection optics, increasing system complexity and cost~\cite{bianconi2025requirements}.

Recently, random modulated continuous-wave (RMCW) LiDAR has emerged as an alternative, modulating laser intensity with a pseudo-random binary sequence to achieve noise-like coding~\cite{yao2025anti}. This approach retains the time-domain measurement principle of pulsed LiDAR while offering improved resistance to interference from other sources~\cite{zhi2025differential}. However, external modulation components introduce additional hardware complexity. Another promising direction is chaotic LiDAR, which utilizes broadband chaotic waveforms generated directly from semiconductor lasers~\cite{chen2023breaking}. By correlating the reflected chaotic signal with a time-delayed reference, centimeter-level ranging resolution can be achieved without external modulation, offering a simpler and more efficient system architecture~\cite{hu2023improving}.

In this work, we demonstrate a high-performance chaotic LiDAR system based on the orthogonal polarization dynamics of a semiconductor VCSEL. VCSELs offer advantages including compact size, low threshold current, low cost, and stable operation, making them highly suitable for practical ranging applications~\cite{liang2024evolution, han2022high}. By introducing delayed orthogonal feedback via an external ring cavity and inserting a half-wave plate to fine-tune the coupling between TE and TM modes, we selectively generate amplified random spikes in the TM mode and fast oscillations in the TE mode. The TM mode serves as a reference, while the TE mode is directed toward the target. The reflected echo is anticorrelated with the reference signal, enabling precise time-of-flight extraction with strong anti-interference capability. Our system achieves linear ranging response and a resolution of centimeter, with further performance tuning possible through adjustment of the half-wave plate angle or laser drive current. Thus, these features collectively enable a robust, tunable, and compact chaotic LiDAR platform, promising enhanced performance in applications requiring high resolution, interference resistance, and system simplicity.

\section{Experimental setup}
\begin{figure}[ht!]
\centering
  \includegraphics[width=8.5cm]{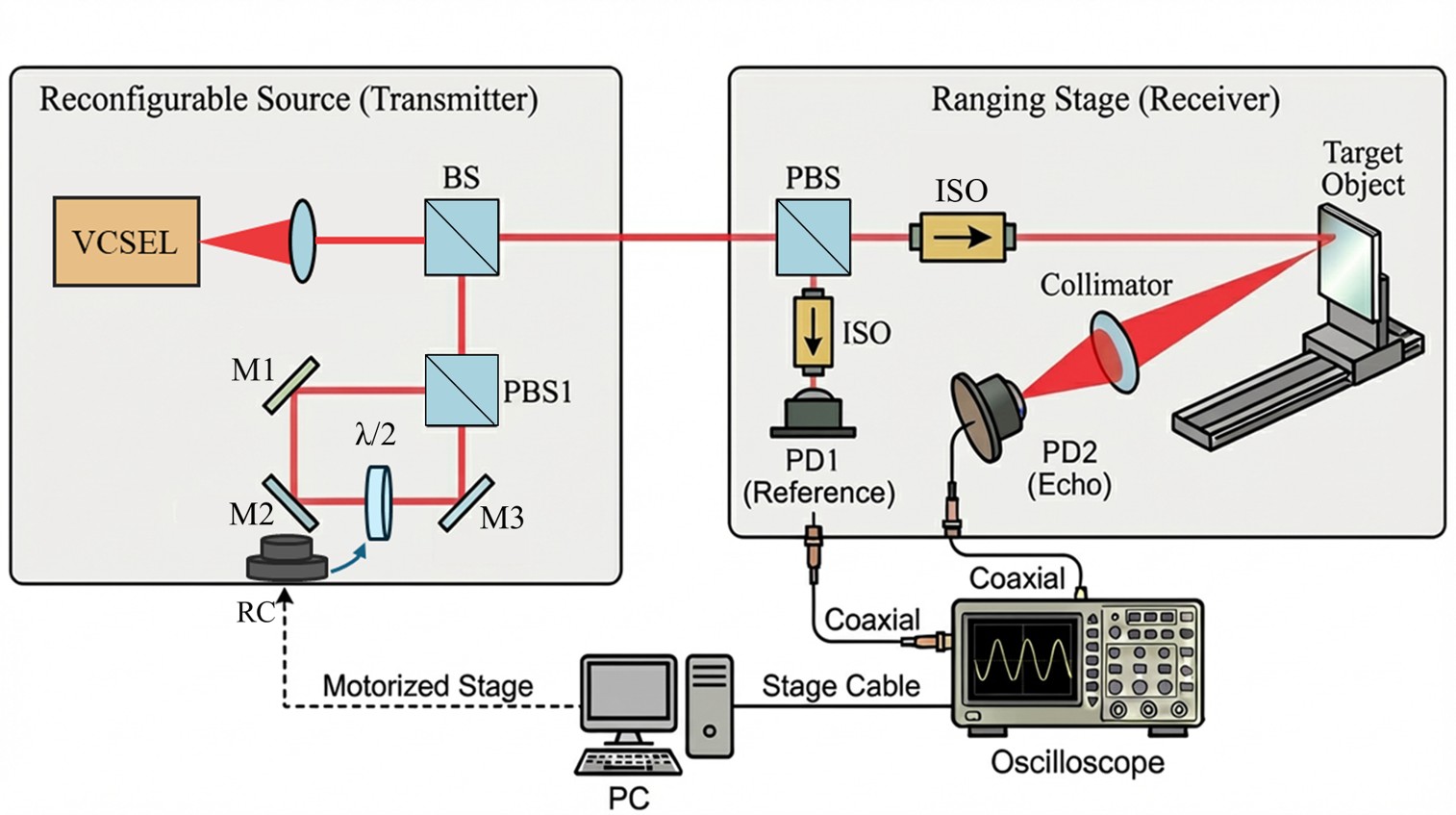}
  \caption{Experimental setup of the chaotic LiDAR system: reconfigurable source (transmitter) and receiver modules. In the transmitter, VCSEL, Vertical Cavity Surface Emitting Laser; BS, beam splitter; PBS1, polarized beam splitter; M1, M2 and M3, mirrors; $\lambda/2$, half-wave plate; RC, rotation controller. In the receiver stage, ISO, optical isolator; PD1 and PD2, photodetectors; PC, personal computer.}
  \label{Experimental_setup}
\end{figure}

The schematic diagram of the LiDAR ranging system is illustrated in Fig.~\ref{Experimental_setup}. The system consists of a transmitter module and a receiver module. In the transmitter, a single-mode semiconductor VCSEL (Thorlabs L850VH1, 850 nm) serves as the light source, with an external ring cavity providing delayed optical feedback. The VCSEL is temperature-stabilized at 25$^\circ$ using a temperature controller (Thorlabs TED200C) and driven by a low-noise current source (Thorlabs LDC205C).  

The collimated laser output first passes through a 50:50 beam splitter (BS), splitting the beam into two branches. The first branch enters the ring cavity, which is constructed using a polarizing beam splitter (PBS1) and three high-reflection mirrors (M1, M2, and M3; OME1-R3-P6, JCOPTIX). Within the ring cavity, a half-wave ($\lambda$/2) plate is inserted to manipulate the polarizations states. The rotation of this plate is controlled via a rotator controller (RC) interfaced with a computer, enabling precise angular adjustment. As light traverses PBS1, the TE and TM polarization modes are separated, and propagating in clockwise and counterclockwise directions, respectively. Consequently, the coupling between these orthogonal modes is dynamically controlled via the angle of the $\lambda$/2 plate~\cite{wang2025mode, wang2025fast}.

The second optical branch is directed to the receiver stage. After passing through a second polarizing beam splitter (PBS2), the TM mode is detected using a DC-coupled amplified photodetector (PD1), serving as the reference signal. The TE mode is transmitted toward a target, and the reflected echo is collected and detected by a second photodetector (PD2). Both signals are recorded simultaneously using a high-bandwidth digital oscilloscope (Tektronix DSA72004, 20 GHz bandwidth). Subsequent cross-correlation analysis is performed computationally to extract the time-of-flight and corresponding target distance. By utilizing the orthogonal polarization states of the VCSEL, the system inherently separates the reference and probe signals without additional optical modulators or complex filtering. Such method can minimize crosstalk, reduces optical loss, and simplifies the detection path.

\section{Results and discussions}
\begin{figure*}[ht!]
\centering
  \includegraphics[width=13cm]{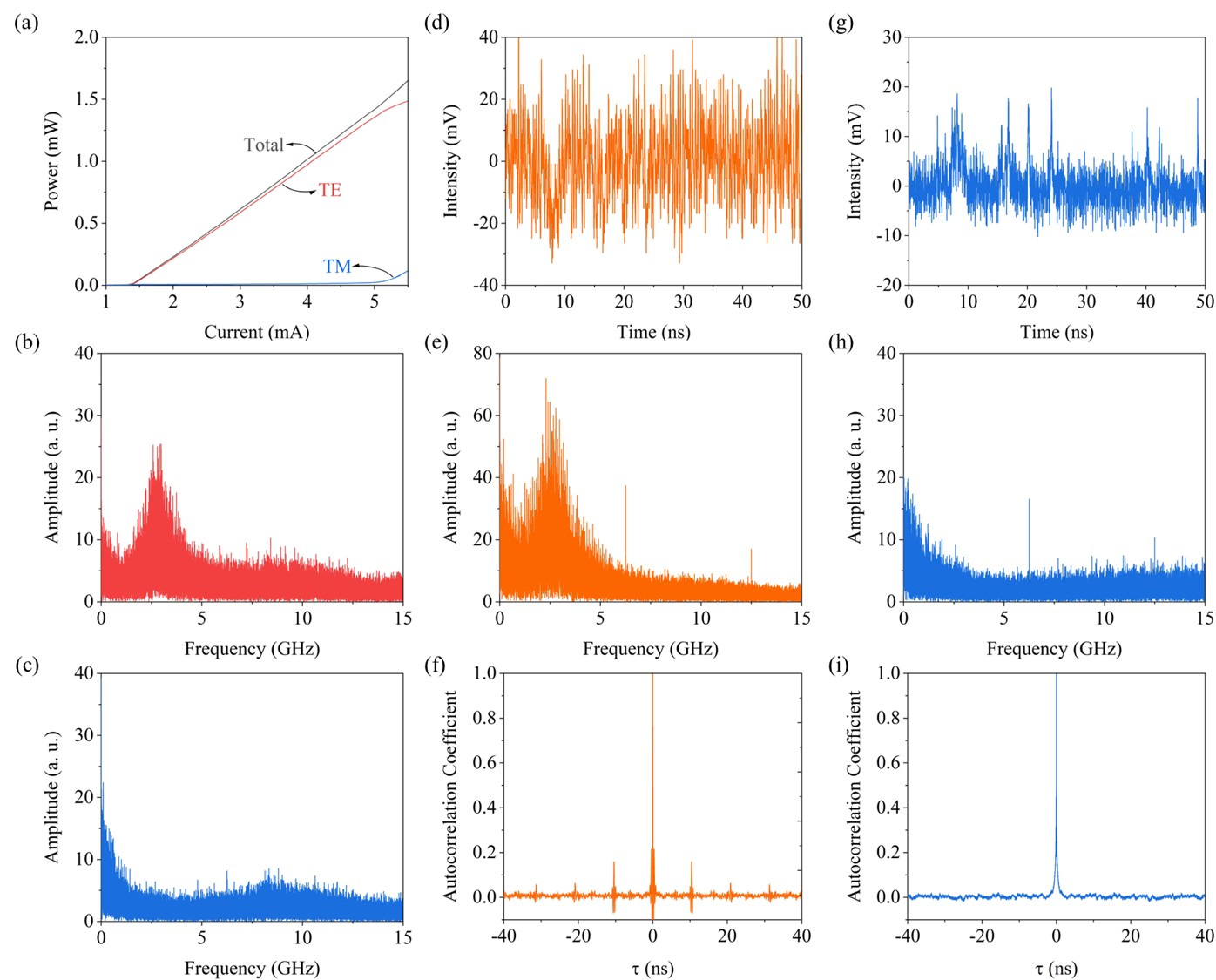}
  \caption{Static and dynamic characterization of the VCSEL under free-running and delayed orthogonal feedback conditions: (a) Light-current (L-I) curves for the total output (black) and the individual TE (red) and TM (blue) polarization modes; (b) and (c) Radio-frequency (RF) spectra of the (b) TE and (c) TM modes for the free-running laser at $J = 1.64$ mA; (d)–(f) Dynamics of the TE mode under orthogonal feedback: (d) temporal trace, (e) corresponding RF spectrum and (f) autocorrelation function. (g)–(i) Dynamics of the TM mode under the same feedback: (g) time series, (h) RF spectrum, and (i) autocorrelation function.}
  \label{Fundamental_characterization}
\end{figure*}

The static and dynamic properties of the vertical-cavity surface-emitting laser (VCSEL) under both free-running and feedback conditions are essential for understanding its operation as a chaotic LiDAR source. Figure~\ref{Fundamental_characterization}a presents the light-current (L-I) characteristics for the total emission (black curve) and the individual transverse electric (TE, red) and transverse magnetic (TM, blue) polarization modes. The TE mode is the dominant lasing polarization, exhibiting a threshold current identical to that of the total output ($\sim$1.6 mA). In contrast, the TM mode remains strongly suppressed at currents below approximately 5.20 mA, operating primarily in the regime of amplified spontaneous emission (ASE) \cite{lindemann2023}. This polarization selectivity is intrinsic to the VCSEL structure and establishes the TE mode as the primary carrier of coherent optical power, while the TM mode provides a sub-threshold, noise-like signal component.

The typical dynamical properties of the free-running laser is examined in Fig.~\ref{Fundamental_characterization}b and c, which display the radio-frequency (RF) spectra of the TE and TM modes at a pump current of $J = 1.64$ mA. The TE mode spectrum (Fig.~\ref{Fundamental_characterization}b) features a pronounced relaxation oscillation peak centred at approximately 2.7 GHz, indicative of its coherent, above-threshold dynamics. Conversely, the TM mode spectrum (Fig.~\ref{Fundamental_characterization}c) is dominated by low-frequency broadband components, consistent with its ASE-driven, incoherent nature below threshold~\cite{frougier2015accurate}. 

Introducing delayed orthogonal-feedback via the external ring cavity dramatically tailors the laser dynamics. The time trace of the TE mode (Fig.~\ref{Fundamental_characterization}d) exhibits fast, irregular oscillations, whose corresponding RF spectrum (Fig.~\ref{Fundamental_characterization}e) shows broadening and intensity enhancement compared to the free-running case. This indicates the degradation of coherence and chaotic modulation induced by the reinjected field. The autocorrelation function of the TE intensity (Fig.~\ref{Fundamental_characterization}f) exhibits a sharp central peak accompanied by symmetric sidelobes, a signature of delayed feedback that imposes a periodic memory on the otherwise stochastic waveform~\cite{wang2021nanolasers, wang2020photon}.

Simultaneously, the TM mode evolves into a train of random, intensity-varying spikes (Fig.~\ref{Fundamental_characterization}g). Its RF spectrum (Fig.~\ref{Fundamental_characterization}h) remains concentrated at low frequencies, confirming that its dynamics are not governed by relaxation oscillations but by amplified stochastic fluctuations. The autocorrelation of the TM mode (Fig.~\ref{Fundamental_characterization}i) displays a rapidly decaying, featureless profile characteristic of broadband chaos, without distinct peaks corresponding to external cavity modes. This confirms that the TM mode operates as an aperiodic, noise-like reference signal. 

\begin{figure}[ht!]
\centering
  \includegraphics[width=8.5cm]{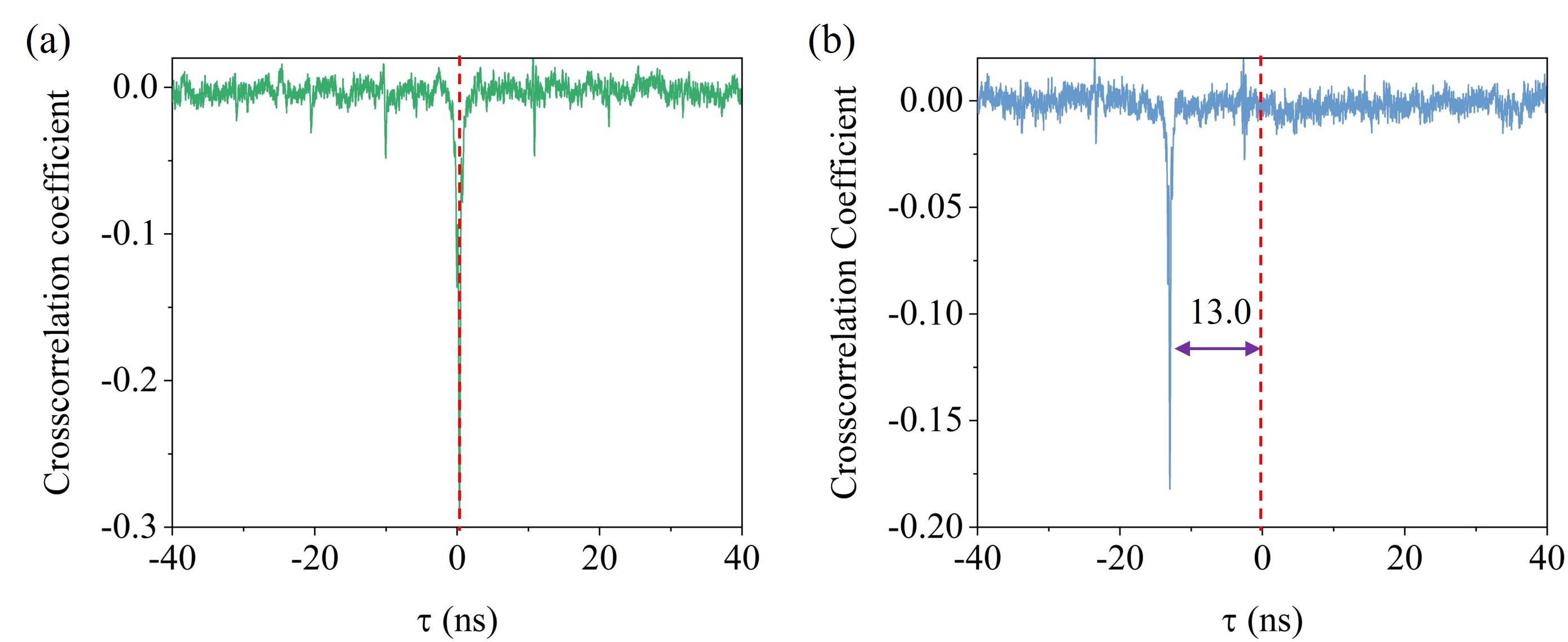}
  \caption{Cross-correlation characterization of the chaotic LiDAR system. (a) CCF between the TE and TM polarization modes under orthogonal feedback. CCF between the TE and TM polarization modes when a target echo is detected.}
  \label{Crosscorrelation_characterization}
\end{figure}

Fig.~\ref{Crosscorrelation_characterization}a presents the cross-correlation function (CCF) between the TE and TM polarization modes under delayed orthogonal feedback. A pronounced inverted (anti-correlation) peak is observed at zero time delay ($\tau = 0$), signifying a negative correlation between the two orthogonal polarization states. This anti-correlation is due to the competitive dynamics induced by the polarization-selective feedback loop, wherein intensity increases in one mode correspond to decreases in the other, corresponding to one typical property of nonlinear mode coupling in the laser cavity~\cite{wang2025mode, ackemann2025polarization}. In addition to the central anti-correlation peak, symmetric sidelobes (wings) are observed, which are a direct signature of the delayed optical feedback from the external ring cavity. These recurring correlation features correspond to the round-trip time of the feedback loop and confirm the periodic reinjection of the orthogonal polarization components into the laser cavity. 

When putting a target into the optical path, the TE mode echo signal reflected from the target is captured by the detector. The resulting CCF shifts distinctly from $\tau = 0$ (Fig.~\ref{Crosscorrelation_characterization}b). The observed time shift of approximately 13 ns corresponds to the round-trip propagation delay between the detector and the target. Using the relation $d = c \cdot \Delta \tau/2$ (where $c$ is the speed of light), this delay translates to a target distance of $\sim$1.95 m. This clear temporal displacement of the anti-correlation peak demonstrates the ranging capability of the system.

The sharpness of the correlation peak and its well-defined shifting under target reflection collectively suggest excellent time-domain resolution and low timing jitter, which are significant for precise distance measurement. Furthermore, the use of an inherently anti-correlated dual-polarization source enhances robustness against common-mode noise and external interference. Therefore, the result validates the proposed chaotic LiDAR scheme,  without the need for high-speed modulation or coherent detection.

\begin{figure*}[ht!]
\centering
  \includegraphics[width=13.5cm]{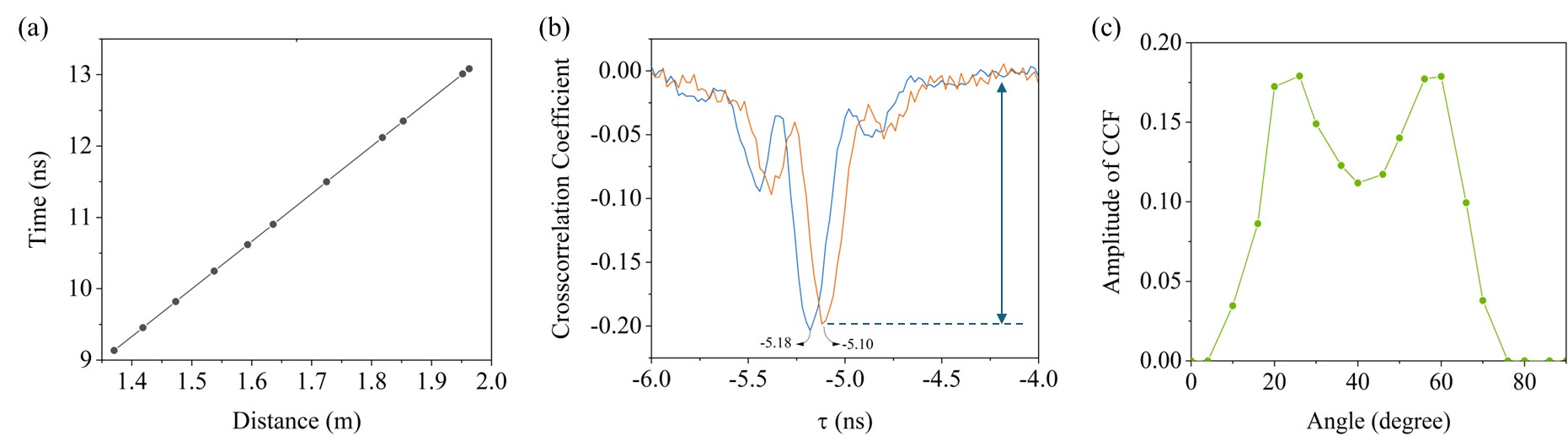}
  \caption{Linearity, resolution, and tunability of the chaotic LiDAR system. (a) Measured time shift of the anti-correlation peak as a function of target distance; (b) Fine displacement measurements; (c) Amplitude of the cross-correlation anti-correlation peak as a function of the $\lambda$/2-plate angle}
  \label{Linearity_resolution}
\end{figure*}

To comprehensively evaluate the ranging performance of the system, we quantified the relationship between the temporal shift of the anti-correlation peak and the physical target distance. Figure~\ref{Linearity_resolution}a plots the measured time shift $\Delta\tau$ as a function of the distance between the target and the detector. The data reveal a precise linear relationship, confirming that $\Delta \tau$ scales directly with the round-trip propagation delay. This linear response validates the fundamental time-of-flight operating principle and ensures reliable distance decoding without complex calibration.

We further characterized the practical resolution of the system by displacing the target in fine increments. As shown in Fig.~\ref{Linearity_resolution}b, the minimum detectable step change in distance is on the order of centimeter, which is consistent with the temporal width of the cross-correlation peak and the effective bandwidth of the chaotic waveform. This resolution is competitive with conventional chaotic LiDAR systems and can be attributed to the broad, noise-like spectrum generated under delayed orthogonal feedback.

One distinctive feature of our setup is the ability to dynamically tune the system's sensitivity via the $\lambda/2$-plate within the feedback loop. As shown in Fig.~\ref{Linearity_resolution}c, rotating the wave plate modifies the polarization coupling between the TE and TM modes, thereby adjusting the amplitude of the anti-correlation peak. This adjustment directly influences the signal-to-noise ratio (SNR) of the correlation signal, enabling real-time optimization of detection sensitivity and ranging accuracy for different operational scenarios. Thus, the $\lambda/2$-plate serves as an in-line gain control that tailors the chaotic dynamics without altering the laser drive conditions or external feedback length, enabling the practicality of our system for robust, reconfigurable optical ranging applications.

\begin{figure}[ht!]
\centering
  \includegraphics[width=8.5cm]{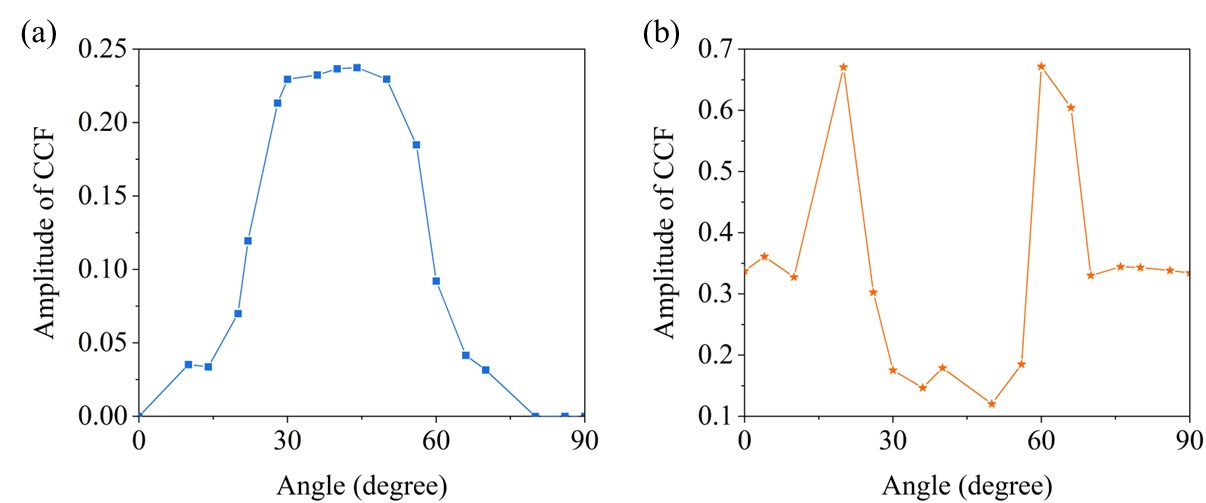}
  \caption{Tunable operating regimes of the chaotic LiDAR system via current and polarization control: (a) $J = 2.10$ mA; (b) $J = 5.76$ mA.}
  \label{SNR} 
\end{figure}

Another distinct advantage is that our chaotic LiDAR system exhibits a broad operational tunability, enabling performance optimization across varying conditions. Fig.~\ref{SNR}a presents this capability by plotting the amplitude of the anti-correlation peak as a function of the $\lambda/2$-plate angle ($\theta$) for two distinct pump currents. At a pump current of $J = 2.10$ mA, the amplitude increases with $\theta$, reaches a plateau of maximum values within the range $30^\circ \leq \theta \leq 50^\circ$, and subsequently declines. This indicates an optimal polarization-coupling regime where the interaction of the TE and TM modes gives the highest signal-to-noise ratio (SNR) for correlation-based ranging. 

When the pump current is elevated to $J = 5.76$ mA, the system enters a different dynamical state. Here, larger anti-correlation amplitudes (exceeding 0.3) are sustained within two distinct angular regions: $\theta \leq 26^\circ$ and $\theta \geq 60^\circ$. This bimodal response demonstrates that the system's operating point can be selectively tuned not only via the$\lambda/2$-plate but also through the injected current, effectively broadening the parameter space for achieving high-performance detection. 

\begin{figure}[ht!]
\centering
  \includegraphics[width=8.5cm]{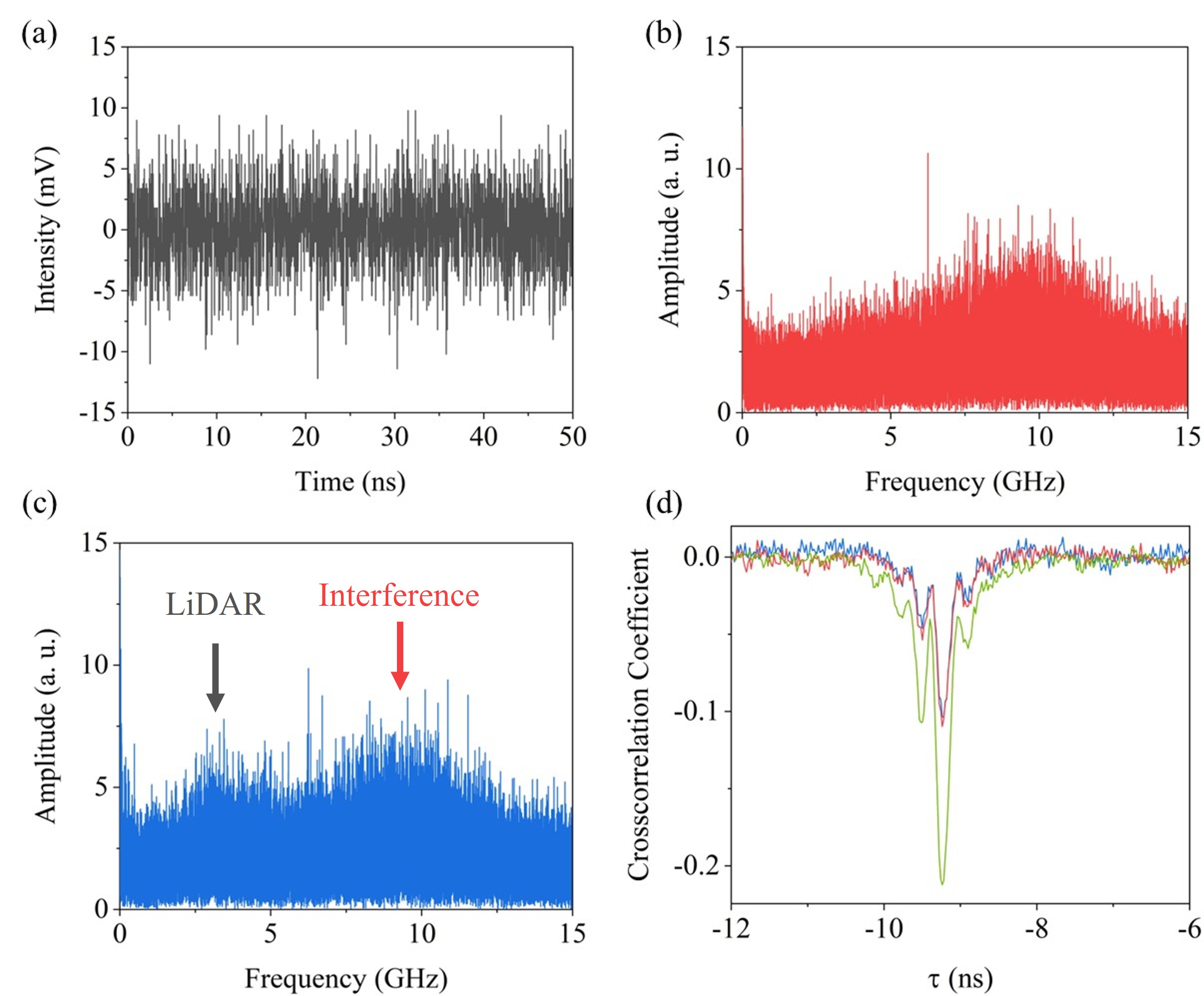}
  \caption{Evaluation of anti-interference capability in a multi-source environment: (a) time trace of an intentional optical interference signal generated by an external 850 nm laser at $J = 73$ mA; (b) Corresponding RF spectrum of the interference; (c) RF spectrum of the combined signal received by the detector; (d) cross-correlation function (CCF) under interference (black) and with varied interference intensity (blue).}
  \label{Anti_jam} 
\end{figure}

Finally, we evaluated the anti-interference capability of our LiDAR system, a critical metric for operation in spectrally crowded environments. To simulate realistic optical crosstalk, an intentional interference signal was generated using a separate 850 nm semiconductor laser (Thorlabs L850P030) operating at a drive current of $J = 73$ mA (far above threshold). The temporal trace of this interfering source, shown in Fig.~\ref{Anti_jam}a, exhibits fast oscillations. Its corresponding RF spectrum (Fig.~\ref{Anti_jam}b) is dominated by a broad peak centered near 10 GHz, characteristic of the relaxation oscillations and intrinsic noise of the free-running laser.

This external interference was optically combined with the echo signal from the target and received by the detector. The RF spectrum of the combined signal (Fig.~\ref{Anti_jam}c) reveals a clear spectral distinction: the chaotic LiDAR signal remains distinct near 2.5 GHz, whereas the interference contributes broadband components that effectively act as additive noise across the detection band.

Notably, despite the presence of this strong interferer, the cross-correlation function (CCF) between the TE and TM channels continues to exhibit a well-defined anti-correlation peak at the correct time delay corresponding to the target distance. Although a minor reduction in the CCF amplitude is observed-likely due to suboptimal alignment under mixed-signal conditions, the temporal position of the peak remains unchanged. This resilience is further demonstrated by varying the intensity of the interference source (blue curve in Fig.~\ref{Anti_jam}d); even as the interference level changes, the anti-correlation peak retains its original time shift, confirming that the ranging information is preserved.

This robust behaviour may origin from two key properties of our system: first, the pseudo-orthogonal and noise-like nature of the chaotic TM reference, which exhibits low cross-correlation with the structured interference; and second, the common-mode rejection inherent in the dual-polarization correlation detection, which suppresses signals not shared by both the TE and TM paths. These results affirm that the proposed chaotic LiDAR architecture maintains accurate ranging functionality even in the presence of significant optical interference, underscoring its suitability for deployment in multi-user or ambient-light-rich environments.   

\section{Conclusions}
In summary, we have successfully demonstrated a novel and highly adaptable chaotic LiDAR system based on the polarization-resolved dynamics of a semiconductor VCSEL with delayed orthogonal feedback. By exploiting the intrinsic TE and TM mode competition, our system generates a pair of complementary optical signals: a chaotic TM reference featured by amplified spontaneous emission and a modulated TE probe with deterministic delayed feedback wings. 

The system exhibits remarkable versatility and flexibility, enabled by two key tuning mechanisms: the half-wave plate for precise polarization coupling adjustment and the pump current for dynamic operation point selection. These controls allow real-time optimization of the cross-correlation amplitude, signal-to-noise ratio, and detection sensitivity across a wide range of operational scenarios. Furthermore, the system maintains linear ranging performance with centimeter-level resolution and demonstrates strong resilience against external optical interference, confirming its suitability for deployment in spectrally congested or multi-user environments.

This innovative polarization-multiplexed approach eliminates the need for external modulators or complex coherent detection, significantly simplifying system architecture while enhancing robustness. Thus, our proposed LiDAR system that is not only high-performing and accurate but also inherently low-cost, energy-efficient, and scalable-attributes that address key barriers to the widespread adoption of advanced ranging technologies. It holds significant promise for applications in autonomous navigation, robotic perception, and secure ranging.

\section*{Acknowledgements}
This work is partially supported by National Natural Science Foundation of China (Grant No. 62475206 and 61804036), and Key Research and Development Plan of Shaanxi Province of China (Grant No. 2024GH-ZDXM-42). 



\bibliography{biblio}

\end{document}